\documentclass{amsart}

\usepackage[foot]{amsaddr}                  
\usepackage{amsmath}                        
\usepackage{amssymb}                        
\usepackage{dsfont} 	                    
\usepackage{bm}                             
\usepackage{mathtools}                      
\usepackage[hypertexnames=false]{hyperref}  
\usepackage[scientific-notation=true]{siunitx}
\usepackage{blkarray}
\usepackage{enumerate}
\usepackage{centernot}
\usepackage{paralist}
\usepackage{stackrel}
\usepackage[table]{xcolor}
\usepackage[colorinlistoftodos,prependcaption,textsize=tiny]{todonotes} 
\usepackage{pagenote}
\usepackage{setspace}
\usepackage[english]{babel}
\usepackage{graphicx}

\usepackage{natbib}

\usepackage{algorithm}
\usepackage{algpseudocode}

\usepackage{listings}
\usepackage{color}

\usepackage{amsthm}
\theoremstyle{plain}
\theoremstyle{definition}

\theoremstyle{remark}

\DeclareMathOperator*{\argmin}{arg\,min}    
\newcommand{\LA}{\mathbf{\Lambda}}

\newcommand{\PS}{\mathbf{\Psi}}
\newcommand{\TA}{\mathbf{\Theta}}

\newcommand{\xih}{\hat{\xi}}

\newcommand{\Y}{\mathbf{Y}}
\newcommand{\Yh}{\hat{\mathbf{Y}}}
\newcommand{\X}{\mathbf{X}}

\newcommand{\y}{\mathbf{y}}

\newcommand{\E}{\mathbf{E}}

\newcommand{\G}{\mathbf{G}}
\newcommand{\Gh}{\hat{\mathbf{G}}}
\newcommand{\g}{\mathbf{g}}
\newcommand{\gh}{\hat{\mathbf{g}}}

\newcommand{\K}{\mathbf{K}}

\newcommand{\I}{\mathbf{I}}

\newcommand{\phic}{\mathbf{\phi}}

\newcommand{\Vh}{\hat{\mathbf{V}}}
\newcommand{\covmat}{\boldsymbol{\Sigma}}
\newcommand{\covmath}{\hat{\boldsymbol{\Sigma}}}
\newcommand{\cormat}{\mathbf{R}}
\newcommand{\cormath}{\hat{\mathbf{R}}}
\newcommand{\Covmat}{\boldsymbol{\Sigma}}
\newcommand{\Covmath}{\hat{\boldsymbol{\Sigma}}}
\newcommand\independent{\protect\mathpalette{\protect\independenT}{\perp}}
\def\independenT#1#2{\mathrel{\rlap{$#1#2$}\mkern2mu{#1#2}}}
\DeclareMathOperator{\vect}{vec}

\makeatletter
\newcommand{\addresseshere}{%
	\enddoc@text\let\enddoc@text\relax
}

\definecolor{lightgrey}{rgb}{0.9,0.9,0.9}
\definecolor{darkgreen}{rgb}{0,0.6,0}
\graphicspath{ {images/} }

\calclayout

\begin{document}

	\title[A latent factor approach to hyperspectral time series data]
	{A latent factor approach to hyperspectral time series data for multivariate genomic prediction \\of grain yield in wheat
	}
	
	\author[J.F.\ Kunst et al.]{Jonathan F.\ Kunst$^1$}
	\address[$^1$]{
		Mathematical \& Statistical Methods group (Biometris) \\
		Wageningen University \& Research \\
		Wageningen\\
		The Netherlands}
	\email{jonathan.kunst@wur.nl}
	
	\author[]{Killian A.C.\ Melsen$^1$}

	\author[]{Willem Kruijer$^1$}

    \author[]{José Crossa$^2$}
    \address[$^2$]{
		Global Wheat Program \\
        International Maize and Wheat Improvement Centre (CIMMYT)\\
		Texcoco\\
		Mexico}
        
	\author[]{Chris Maliepaard$^3$}
	\address[$^3$]{
		Plant Breeding \\
		Wageningen University \& Research \\
		Wageningen\\
		The Netherlands}
	
	\author[]{Fred A.\ van Eeuwijk$^1$}

	\author[]{Carel F.W.\ Peeters$^1$}


    \begin{abstract}
		\label{abstract}
        \textbf{Key message: We develop and apply a two-stage factor analytic approach with Procrustes rotation for the integration of time series secondary phenotypes into multivariate genomic prediction.}
		High-dimensional time series phenotypic data is becoming increasingly common within plant breeding programmes. 
		However, analysing and integrating such data for genetic analysis and genomic prediction remains difficult. 
		Here we show how factor analysis with Procrustes rotation on the genetic correlation matrix of hyperspectral secondary phenotype data can help in extracting relevant features for within-trial prediction. 
		We use a subset of Centro Internacional de Mejoramiento de Ma\'{i}z y Trigo (CIMMYT) elite yield wheat trial of 2014-2015, consisting of 1,033 genotypes. These were measured across three irrigation treatments at several timepoints during the season, using manned airplane flights with hyperspectral sensors capturing 62 bands in the spectrum of 385-850 nm.
		We perform multivariate genomic prediction using latent variables to improve within-trial genomic predictive ability (PA) of wheat grain yield within three distinct watering treatments.
		By integrating latent variables of the hyperspectral data in a multivariate genomic prediction model, we are able to achieve an absolute gain of .1 to .3 (on the correlation scale) in PA compared to univariate genomic prediction. 
		Furthermore, we show which timepoints within a trial are important and how these relate to plant growth stages. 
		This paper showcases how domain knowledge and data-driven approaches can be combined to increase PA and gain new insights from sensor data of high-throughput phenotyping platforms.
		
	\end{abstract}
	
	\maketitle


	\section{Introduction}\label{SEC:Intro}
	Genomic prediction has aided plant breeding over the last decade in a variety of ways.
	By capturing a large number of small additive genetic effects, prediction of complex polygenic traits (e.g., yield) improved for those crops which have a dense marker set available \citep{jannink_genomic_2010}.
    It has reduced the cost and timepath of genetic gain relative to conventional phenotypic methods \citep{wartha_implementation_2021}.
	In addition, the flexibility of genomic prediction methods has advanced parental selection and multi-environmental evaluation within plant breeding programmes \citep{jarquin_genomic_2020, lopez-cruz_optimal_2021}.

	While univariate prediction is relatively straightforward, multivariate genomic prediction is more suitable when additional secondary phenotype information is available. 
	Secondary phenotype information could of course be added as predictors in a univariate prediction model. 
	But from a theoretical perspective, a multivariate model is more correct, as genetic effects have influence over both secondary traits and focal traits. 
	In essence, highly heritable secondary traits can aid in prediction of a less heritable complex trait when a strong genetic correlation exists between traits \citep{thompson_review_1986}. 
	For example, multi-trait genomic prediction can be used for multiple disease characteristics \citep[e.g., presence, absence and shape,][]{Jia2012} or grain quality traits and yield \citep{Michel2019}.
	Accordingly, integrating many traits in a genomic prediction model has been an active field of research in the past decade \citep{krause_hyperspectral_2019, dos_santos_novel_2020, arouisse_improving_2021, runcie_megalmm_2021, melsen_2025}.
	
	High-throughput phenotyping technologies have been developed in parallel with genomic prediction methods. 
	These technologies help evaluate genotypes in numerous ways, from cell to organism levels \citep{van_dijk_machine_2021}.
	For instance, unmanned aerial vehicles (UAVs) equipped with sensors have made evaluation of field trials across a season more feasible. 
	Meanwhile, controlled environment facilities with automated imaging and treatment application can provide extensive information on individual plant treatment responses.
	These advancements have resulted in tremendous amounts of phenotyping data generated by each experiment.
	
	Using high-throughput phenotyping data in multivariate genomic prediction is complicated due to its high-dimensional nature.
	Fitting a model with many traits may run into problems due to singularity and multicollinearity when using an unstructured covariance matrix for residuals between traits \citep{antonion_multivariate_2022}.
	Consequently, in practice some form of dimensionality reduction or regularization is required. 
	For example, by regressing the focal trait onto the secondary phenotype data through ridge or LASSO regression and subsequently using the prediction in a bivariate prediction model \citep{lopez-cruz_regularized_2020, arouisse_improving_2021}.
    While increasing predictive ability (PA) is a goal in itself, interpretability of methods is an important aspect to keep in mind. 
	We often want to know how secondary trait information relates to a complex genetic trait and what aspects of this relation are important for improving prediction of a focal trait.
    Moreover, knowing which traits and timepoints are important may increase efficiency of future phenotyping efforts, as plant breeders can be more selective in traits measured and moments of phenotyping.    
    While regularization methods provide interpretable models or coefficients, they are often based on untenable sparsity assumptions and may lead to unstable solutions \citep{Giannone_2021}.
    Dimension reduction methods project the original features onto a lower-dimensional representation of the data.
    Such projections do not have to be amenable to a sparsity assumption and may avoid overpenalization \citep{Manthena_2022} with the potential caveat of reduced interpretability. 
    Here, we provide an interpretable dimension reduction method for high-dimensional secondary phenotypes over time to support multivariate genomic prediction.

	Factor analysis is a widely used dimensionality reduction technique in many fields and it has gained traction within plant breeding as well. 
	Already appearing as a way to deal with large genotype-by-environment matrices, it is now utilized in reducing the dimensionality of phenotype datasets and to summarize correlated traits  \citep{piepho_empirical_1998, runcie_megalmm_2021, paixao_factor_2022, fialho_factor_2023, melsen_2025}.  
	The construction of latent factors can be unsupervised and based on the genetic correlation matrix of secondary high-dimensional traits such as hyperspectral data.
	In addition, these latent factors are available for further biological interpretation through rotational techniques.
	For example, through Procrustes rotation \citep{schonemann_generalized_1966} we are able to show how the representation of secondary traits is relatively stable across time by aligning the factor structures of different timepoints.
    In short, factor structures of all timepoints are rotated towards the factor structure of a timepoint of choice, where the timepoint of choice can be based on domain knowledge or data-driven. 
    Here domain knowledge of wheat can guide selection, where we choose the timepoint of heading as anchor to rotate other timepoints towards, as heading date has a large impact on grain yield \citep{Araus_2008}.

	In this paper we present the application of factor analysis with Procrustes rotation for interpretation and integration of high-dimensional time series phenotype data into genomic prediction models. 
	We show how this method can be applied to hyperspectral data collected for three elite yield trials of the Centro Internacional de Mejoramiento de Ma\'{i}z y Trigo (CIMMYT) with different watering regimes \citep{krause_hyperspectral_2019}.
	By incremental modeling steps we illustrate how secondary trait information may help in increasing PA and also when phenotyping is most crucial. 
    We thus offer a computationally efficient way of fitting a dynamic factor model while helping breeding with a more efficient phenotyping strategy by showing which timepoints and traits are important.
    

	\section{Materials and Methods}    
\begin{figure}[ht]
		\centering
		\includegraphics[width=\textwidth]{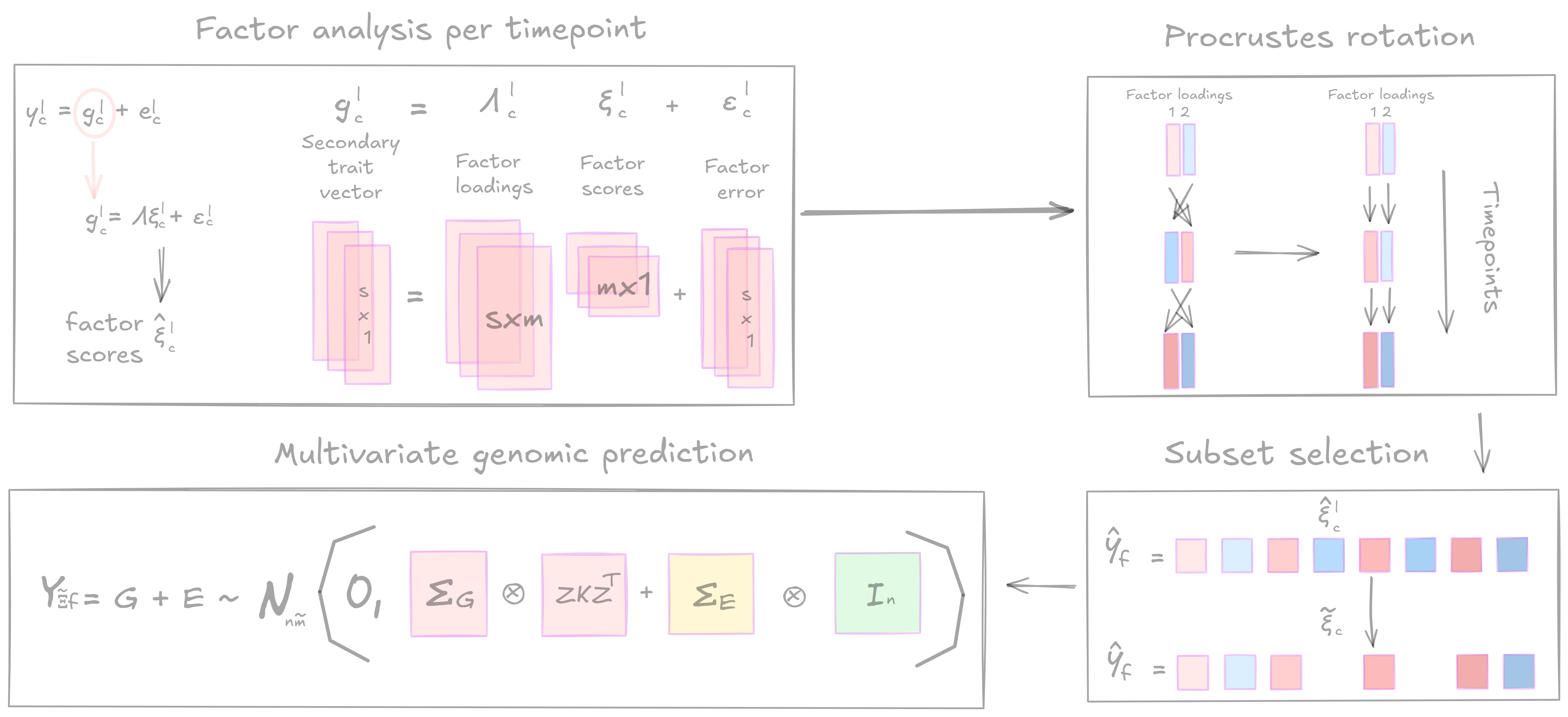}
		\caption{
			A graphical illustration of our dimensionality reduction and multivariate genomic prediction workflow. Starting with a genetic covariance matrix of secondary traits (red), we estimate lower dimensional factor scores and loadings per timepoint. The factor loadings are reordered using Procrustes rotation to trace factor scores across timepoints. Next, we use subset selection to relate factor scores ($\mathbb{\hat{\xi}}$) to the focal trait ($\mathbf{\hat{\y}}_{f}$). The selected factor scores ($\mathbf{\tilde{\Xi}}$) and focal trait ($\mathbf{\hat{\y}}_{f}$) are then used in multivariate genomic prediction for unseen genotypes through a (genomic) relationship matrix ($\mathbf{K}$). Complete explication of the notation and modeling can be found in the sections below.}
		\label{graphical_abstract}
	\end{figure}

    We propose a two-stage approach to multivariate genomic prediction with temporal, high-dimensional secondary phenotypes. 
    We first use a specific rotational approach to dynamic factor analysis to produce a temporally consistent lower-dimensional representation (factor scores of latent factors) of our secondary features.
    The low-dimensional representations are subsequently used as additional traits in multivariate genomic prediction.
    Our approach is graphically represented in Figure \ref{graphical_abstract}.
    Below, we will introduce the motivating data and our notation before explicating our two modeling stages.

	\subsection{Motivating data: CIMMYT elite yield trial}
	We use the CIMMYT elite wheat yield trial dataset generated at the Norman E.\ Borlaug Research Station in Ciudad Obreg\'{o}n, M\'{e}xico \citep{krause_hyperspectral_2019}. 
	In this dataset, the secondary traits are hyperspectral wavelength reflectivities and the focal trait is grain yield.
	We focus on a subset consisting of trials from 2014--2015, characterised by three different irrigation treatments of increasing drought: optimal flat ($5$ furrow irrigations), moderate drought ($3$ furrow irrigations) and severe drought ($2$ furrow irrigations). 
    Within these treatments, $1,033$ genotypes, each with three replicates, were evaluated with hyperspectral data collected using an aircraft across several timepoints ($7$ to $10$ dates, depending on treatments). 
    The spectrum range is $385-850$ nm, with an interval of $7.5$ nm, resulting in reflectivities of $62$ wavebands measured at each timepoint.
	The reflectivities were based on the average of the middle of the plots within a trial.
	In addition, the population was genotyped using $8,519$ single nucleotide polymorphism (SNP) markers.
	For more details see \citet{krause_hyperspectral_2019}.
	
	\subsection{Notation} 
    We use phenotype data of which the experimental design has been resolved, as done in \citet{lopez-cruz_regularized_2020}.
	Of genotypes $c = 1,...,g$ we have $j = 1,..., s$ secondary traits of $i = 1, ...,n$ plots with replicates $q = 1,...,r$ and timepoints $l = 1,..., \tau$. 
    In our data context we are considering a balanced design (although this is by no means a necessity for the methodology proposed below) with $g = 1,033$, $s = 62$, and $r = 3$.
    This results in $n= gr = 3,099$ plots and, depending on the treatment, $\tau = 10$ for optimal flat, $\tau = 8$ for moderate drought, and, $\tau = 7$ for severe drought. 
    
    For vectors we use boldface lowercase symbols (i.e, $\bm{a}$ for a vector) and for matrices boldface uppercase symbols (i.e, $\mathbf{A}$ for a matrix). 
    When relevant, we specify the row of a data matrix such that $\boldsymbol{a}_{(q|c)}$  represents the data vector for replicate $q$ given genotype $c$.
    We will usually denote the kinship matrix by $\mathbf{K}$.
    Also, we use $\mathbf{\Sigma}$ and $\mathbf{R}$ to denote, respectively, covariance and correlation matrices.
    We use the subscripts $P$, $G$, and $E$ to denote phenotypic, genetic, and residual covariance and correlation matrices.
    For example, $\mathbf{\Sigma}_{E}$ denotes a residual covariance matrix.
    The superscipt $l$ refers to the timepoint.
    Then, for example, $\mathbf{\Sigma}_{E}^{l}$ denotes a residual covariance matrix for timepoint $l$.
    We use $\mathbf{I}_n$ to denote the $(n \times n)-$dimensional identity matrix.

    We use, in terms of operations, $\otimes$ for the Kronecker product and $\circ$ for the Hadamard product.
    We let $\mathrm{diag}(a_{11}, \ldots, a_{ss})$ denote an $(s \times s)$-dimensional diagonal matrix with dagonal elements $a_{11}, \ldots, a_{ss}$. 
    The column-wise vectorization of a matrix is denoted by $\mathrm{vec}(\mathbf{A})$ and $\mathrm{vec}^{-1}(\mathbf{A})$ denotes the reverse operation.
    Also, if $\mathbf{A}$ is square, we let $\mathbf{A}^{1/2}$ denote the matrix square root and $\mathbf{A}^{-1/2}$ the square root of the inverse $\mathbf{A}^{-1}$.
    In addition, we let $I(\cdot)$ denote an indicator function with argument $(\cdot)$.
    Lastly, the Frobenius norm is represented by $\|\cdot\|_\mathcal{F}$.
		
	\subsection{Projecting high-dimensional secondary traits}
    In the following sections the projection of high-dimensional traits onto lower-dimensional latent variables is described. 
	First, we describe how phenotypic data can be seen as a combination of genetic and residual effects.
    We then estimate the genetic covariance of secondary traits for each timepoint, followed by estimating factor loadings and retrieving factor scores for each timepoint.
	Subsequently, we use Procrustes rotation and smoothing to trace factor scores across timepoints.
	
	\subsubsection{A basic model for secondary traits}
    We describe the complete secondary phenotype matrix per treatment as $\Y \in \mathbb{R}^{n \times s\tau}$. 
    The secondary phenotype matrix partitioned per timepoint is indicated as $\Y^{l} \in \mathbb{R}^{n \times s}$.
	The matrix $\Y^{l}$ is then assumed to be decomposable into genetic ($\G^l$) and residual ($\E^l$) effects, in the sense of $\Y^l$ being a sum of genetic and residual effects. Then, $\vect(\Y^l)$ has the following multivariate normal distribution:
	\begin{equation*}
		\vect(\Y^l) = \vect(\G^l) + \vect(\E^l) \sim \mathcal{N}_{ns}\big(\mathbf{0}, \Covmat_{G}^l \otimes \mathbf{Z}\K\mathbf{Z}^{\top} + \Covmat_{E}^l \otimes \I_{n}\big) \text{,}
	\end{equation*}
	with $\Covmat_{G}^l \in \mathbb{R}^{s \times s}$ and $\Covmat_{E}^l \in \mathbb{R}^{s \times s}$ being the genetic and residual covariance matrices, $\mathbf{Z} \in \mathbb{R}^{n \times g}$  denotes the incidence matrix, and $\K \in \mathbb{R}^{g \times g}$.
    We obtain $\mathbf{\Sigma}^l_G$  through the covariance matrix of the observed best linear unbiased estimators (BLUEs) of the observed (e.g., training set) genotypes and $\mathbf{\Sigma}^l_E$ and from the covariance matrix of the plot level residuals:
	\begin{equation*}
		\mathbf{\bar{\y}}^l_{c} = \mathbf{g}^l_{c} + \boldsymbol{\bar{\epsilon}}^l_{c} = \mathbf{g}^l_c + \frac{1}{r} \sum^r_{q = 1}  \boldsymbol{\epsilon}_{\left(q|c\right)} \text{.}
	\end{equation*}
	The plot level residuals $\boldsymbol{\epsilon}^l_{\left(q|c\right)}$ of genotype $c$ and replicate $q$ are assumed to be independently and identically distributed, so the residual covariance matrix can be shown to be $r^{-1}\covmat^l_{E}$ \citep{kruijer_2020}, which results in $\mathbf{\bar{\y}}^l_{c} \sim \mathcal{N}_{s}\left(\mathbf{0}, \covmat^l_{P}\right)$, where $\covmat^l_{P} = \covmat^l_{G} +  r^{-1}\covmat^l_{E}$.
    The covariance matrices are estimated as described in \citet{melsen_2025}.
    Now we have to focus on $\mathbf{\Sigma}^l_G$ and $\mathbf{\Sigma}^l_E$ for each timepoint, of which the former is used in subsequent factor analysis and the latter is required in multivariate BLUP equations together with the latent factor scores.
	
	\subsubsection{A factor model for the genetic effects}
	We perform factor analysis based on the genetic correlation matrix of spatially corrected genotypic means of all observed secondary traits at each timepoint.
	The genetic component of the secondary traits is factored as:
	\begin{equation*}
		\g^l_c = \LA^l\bm{\xi}^l_c + 
        \boldsymbol{\varepsilon}^l_c \text{,}
	\end{equation*}
	with $\LA^l \in \mathbb{R}^{s \times m}$ being the factor loadings matrix of $s$ secondary traits on $m$ factors, $\bm{\xi}^l_c \in \mathbb{R}^{m\times 1}$ the latent factor of genotype $c$ and, $\epsilon$ the factor model error.
	Here we make the following assumptions \citep{peeters_stable_2019, melsen_2025}: (i) rank$\left(\LA^l\right) = m < s$, (ii) $\bm{\varepsilon}^l_c \sim \mathcal{N}_{s}\left(\mathbf{0}, \mathbf{\Psi}^l\right)$, where $\mathbf{\Psi}^l = \mathbb{E}\left(\boldsymbol{\varepsilon}^l_c \boldsymbol{\varepsilon}^{l \top}_c\right)$ and $\bm{\Psi}^l \equiv \mathrm{diag}\left(\psi^l_{11}, . . .,\psi^l_{ss}\right)$ with $\psi^l_{jj} > 0$, $\forall jj$, (iii) $\bm{\xi}^l_c \sim \mathcal{N}_m \left(\bm{0}, \bm{\I}_m\right)$, (iv) $\bm{\xi}^l_c \independent \boldsymbol{\varepsilon}^l_{c'}, \forall c,c'$, (v) $\bm{\xi}^l_c \independent \bm{\bar{\epsilon}}^l_{c'}, \forall c, c'$, and (vi) $\boldsymbol{\varepsilon}^l_c \independent \bm{\bar{\epsilon}}^l_{c'}, \forall c, c'$.
	Given these assumptions, we decompose the genetic covariance matrix of each timepoint ($\covmat_{G}^l$) as:
	\begin{equation*}
		\covmat_{G}^l\left(\TA^l\right) =
		\mathbb{E}\left[\big(\LA^l\bm{\xi}^l_c + \boldsymbol{\varepsilon}^l_c\big)\big(\LA^l\bm{\xi}^l_c + \boldsymbol{\varepsilon}^l_c\big)^\top\right] =
		 \LA^l\LA^{l\top} + \PS^l
		 \text{,}
	\end{equation*}
	with $\TA^l = \lbrace\LA^l,\PS^l\rbrace$ being the model-implied loadings and uniqueness matrix at timepoint $l$. 
	We estimate the loadings matrix and subsequent factor scores based on the genetic correlation matrix at each timepoint, as standardized loadings between $[-1,1]$ are easily interpretable.
    However, a covariance matrix may also be used as the estimation procedure is scale free, but it will be less interpretable.
	Given the estimated genetic covariance matrices $\covmath_{G}^l$ we scale these to correlation matrices:
	\begin{equation*}\label{eq:covtocor}
		\begin{split}
			\cormath_{G}^l&=\big(\covmath_{G}^l\circ \I_{s}\big)^{-1/2} \covmath_{G}^l \big(\covmath_{G}^l\circ \I_{s}\big)^{-1/2}
		\end{split}.
	\end{equation*}

	\subsubsection{Parameter estimation per timepoint}
	When performing factor analysis the number of factors $m$ (i.e., the factor dimension) needs to be determined.
    We do so by using proxy $m^*$ for $m$. 
    Here we use a numerical alternative to scree-plot elbow detection, that is the acceleration factor \citep{raiche_non-graphical_2013}.
    This method finds the eigenvalue at the point preceding high acceleration, with the lower bound according to Kaiser's rule (i.e., eigenvalues $> 1$, see appendix \ref{sup:dimension}).
    This follows the idea of information diminishing after the elbow point.
	For each timepoint $l$ we estimate the collection of factor model parameters $\hat{\mathbf{\TA}}^l_{m^*} = \{\hat{\LA}^l, \hat{\mathbf{\Psi}}^l\}_{m^*}$  by minimizing the following discrepancy function \citep{joreskog_1967}:
	\begin{equation*}
		F\left[\cormath^l_{G}\left(\TA \right)^l;\cormath_{G}^l;m^*\right] = 
		\text{ln}\left|\LA^l \LA^{l \top} + \mathbf{\Psi}^l\right| + \text{tr}\left[\cormath^l_{G}\left(\LA^l \LA^{l \top} + \mathbf{\Psi}^l\right)^{-1}\right] -
		\text{ln}|\cormath_{G}^l| - s \text{,}
	\end{equation*}
	which requires $\cormath_{G}$ to be a positive definite and well-conditioned sample matrix.
	This results in:
	\begin{equation*}
		\hat{\TA}^l_{m^*} \equiv \argmin_{\LA^l, \Psi^l}  F\left[\cormat^l_{G}\left(\TA \right);\cormath^l_{G},m^*\right].
	\end{equation*}
	 Furthermore, we apply Varimax rotation on the loadings matrix \citep{kaiser_varimax_1958} as it provides an approximately sparse loadings matrix accommodating interpretability. 
     As such, this rotation will let secondary traits have high loadings on one particular factor while retaining lower loadings on other factors.
	
	\subsubsection{Procrustes rotation and smoothing for consistency across timepoints}
	After estimating loadings matrix $\hat{\LA}^l$ for each timepoint $l$, we want to track factor scores across time. 
	As the Varimax rotation criterion is invariant over sign- and label-switching we can lose interpretability of loadings and scores if the rotation per timepoint is solely based on unconstrained Varimax.
    Basically, we want to find, for each time-dependent loadings matrix, a signed permutation matrix that will put all such matrices in the same reflection (i.e., the same ordering of columns and all columns in the same polarity).
    We approach this issue with Procrustes rotation.
    With Procrustes rotation we can rotate each loadings matrix towards a selected target matrix $\hat{\LA}^{l \prime}$ \citep{schonemann_generalized_1966}.
    Consequently, we can represent all time-dependent loadings matrices and factor scores in the same reflection as a biologically relevant target timepoint (e.g., heading date).
	As a result, we can observe dynamics of factors across the season within a trial. 
    In practice, this means finding an orthogonal rotation matrix $\mathbf{T}$ that minimizes the difference between loadings matrices in a least-squares sense \citep{schonemann_generalized_1966}:
    \begin{equation*}
		\mathbf{T}^{l} \equiv \argmin_\mathbf{H} \Big\{\big\|\hat{\LA}^{l }\mathbf{H} - \hat{\LA}^{l \prime}\big\|_\mathcal{F} : \mathbf{H}^\top \mathbf{H} = \I_{m^*}\Big\}.
	\end{equation*}	
    Note that the Procrustean rotation is a strong rotation criterion, in the sense that it may remove part of the (sampling) variability beyond mere permutations in the ordering and polarity of the columns of $\hat{\LA}^{l}$.
    However, $\mathbf{T}^{l}$ will be close to a signed permutation matrix ${_\pm}\mathbf{P}^{l}$.
    We may extract the signed permutation matrix ${_\pm}\mathbf{P}^{l}$ contained in $\mathbf{T}^{l}$ using a smoothing algorithm described in \citet[][Appendix 2A]{Peeters2012}.
	The product $\hat{\LA}^l {_\pm}\mathbf{P}^{l}$ then assures that $\hat{\LA}^l$ is in the same reflection as the target $\hat{\LA}^{l \prime}$. 
    Now we can represent secondary traits through an approximately sparse factor representation, while tracing these across time.

	\subsubsection{Estimating factor scores per timepoint}
\begin{figure}[ht]
		\centering
		\includegraphics[width=\textwidth]{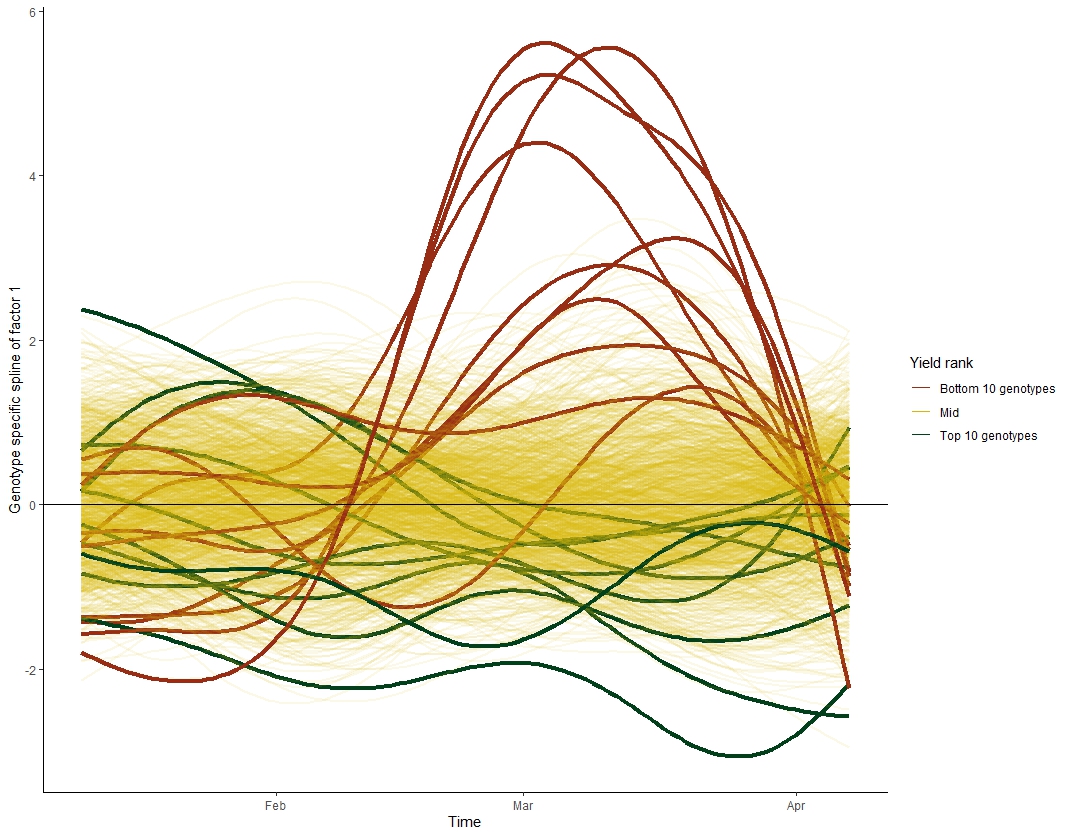}
		\caption{Illustration of using smoothing splines per genotype based on factor 1 within the Optimal Flat treatment. In terms of yield, the bottom 10 genotypes (red) diverge from the best 10 genotypes (green) around March. Moreover, the population spline is indicated by the black line at $y = 0$.
        Hence, one can see the factor space also as a normed space.}
		\label{fig:factor_spline}
	\end{figure}

	The (Procrustean) rotated factor scores for each timepoint, $\hat{\bm{\Xi}}^l$, are estimated through a modified Thomson regression approach \citep{thomson_factorial_1939, melsen_2025}. 
	In short, we project the secondary phenotypes onto the lower dimensional factor space, resulting in:
	\begin{equation*}
			\hat{\mathbf{\Xi}}^l = \bar{\Y}^l\left(
			\hat{\mathbf{\Psi}}^{l} + r^{-1}\covmath_{E}^{l}
			\right)^{-1}
			\hat{\LA}^{l}
			\left[\I_{m^*} + \hat{\LA}^{l\top}
			\left(\hat{\mathbf{\Psi}}^l + r^{-1}\covmath_{E}^{l}
			\right)^{-1}
			\hat{\LA}^{l}
			\right]^{-1}
		\text{.}
	\end{equation*}
	These factor scores can then be directly used in subsequent modeling steps.
	In another way, these factor scores can be traced over time, opening multiple options for modelling.
	One such option is the visualisation of trends across time using smoothing splines \citep{perez-valencia_two-stage_2022, boer_tensor_2023}, which can show relations of factors with trait data, such as yield (Figure \ref{fig:factor_spline}). 
	In addition, if desired, factor scores can be substituted by spline characteristics as shown in appendix \ref{sup:spline parameters}.

	\subsection{Multivariate genomic prediction}
	After obtaining factor scores for each timepoint, we include the best subset of secondary trait factors based on minimizing the Bayesian Information Criterion \citep{Schwarz_1978}. 
    We find the best subset within the training set by regressing the mean focal trait $\mathbf{y}_f$ on the factor scores.
    We start with $\tau m^*$ secondary trait factor scores.	
	After subset selection, we are left with candidate factors $\tilde{m} \leq \tau m^*$, which form the matrix $\mathbf{\tilde{\Xi}}$.
    Subsequently, we combine the selected factor scores and focal trait as:
    \begin{equation*}
        \Y_{\tilde{\Xi} f} =  \begin{bmatrix}
            \mathbf{\tilde{\Xi}} \quad \y_f
        \end{bmatrix},
    \end{equation*}
    with $\mathbf{\tilde{\Xi}}$ being an $n \times \tilde{m}$ matrix containing plot level factor scores and $\y_f$ being the $n \times 1$ column vector of the focal trait.
    This matrix again consists of genetic and residual components as:
    \begin{equation*}
        \Y_{\tilde{\Xi} f} = \mathbf{G}_{\tilde{\Xi} f} + \mathbf{E}_{\tilde{\Xi} f}
    \end{equation*}
    Assuming that the genetic components are independent of the residual components we get:
	\begin{equation*}
		\vect(
		\Y_{\tilde{\Xi} f}
		) =
		\vect\big(\mathbf{G}_{\tilde{\Xi} f}\big) + \vect\big(\mathbf{E}_{\tilde{\Xi} f}\big) \sim \mathcal{N}_{n\tilde{m}}\big(\mathbf{0}, \Covmat_{G} \otimes \mathbf{Z}\K\mathbf{Z}^{\top} + \Covmat_{E} \otimes \I_{n}\big) \text{,}
	\end{equation*} 
    where $\mathbf{Z}\K\mathbf{Z}^{\top}$ and $\I_{n}$ are the row covariances of $\mathbf{G}_{\tilde{\Xi} f}$ and $\mathbf{E}_{\tilde{\Xi} f}$, respectively. Similarly, $\Covmat_{G}$ and $\Covmat_{E}$ are the column covariances of $\mathbf{G}_{\tilde{\Xi} f}$ and $\mathbf{E}_{\tilde{\Xi} f}$.
    These covariance matrices are block-partitioned:
		\begin{equation*}
		\Covmat_{G} = \begin{bmatrix}
			\Covmat^{\tilde{\Xi}}_G  \quad \boldsymbol{\sigma}^{\tilde{\Xi} f}_G\\
			\boldsymbol{\sigma}^{f \tilde{\Xi}}_G \quad  \sigma^{f}_G
		\end{bmatrix}, \quad
		\Covmat_{E} = \begin{bmatrix}
			\Covmat^{\tilde{\Xi}}_E  \quad \boldsymbol{\sigma}^{\tilde{\Xi} f}_E\\
			\boldsymbol{\sigma}^{f \tilde{\Xi}}_E \quad \sigma^{f}_E
		\end{bmatrix} \text{,}
	\end{equation*} 
    where $\Covmat^{\tilde{\Xi}}_G$ and
    $\Covmat^{\tilde{\Xi}}_E$ are the genetic and residual covariances of the factor scores, and where $\sigma^{f}_G$ and $\sigma^{f}_E$ denote the focal trait genetic and residual scalar variances.
    The off-diagonal elements are the covariances of the factor scores and focal trait, with $\boldsymbol{\sigma}^{\tilde{\Xi} f}_G  = \big(\boldsymbol{\sigma}^{f \tilde{\Xi}}_G\big)^\top$ and $\boldsymbol{\sigma}^{\tilde{\Xi} f}_E = \big(\boldsymbol{\sigma}^{f \tilde{\Xi}}_E\big)^\top$. 
    Note that without dimension reduction, these covariance matrices would be much larger in size rendering multivariate genomic prediction computationally expensive.
    This approach to multivariate genomic prediction implies, as the kinship matrix is marker-based, an integration of secondary trait and genomic data.
    See Appendix \ref{sup:blup_equations} for the full multivariate BLUP equations for cross-validation scenarios 1 (secondary data only available for the training set) and 2 (secondary data available for both training and testing set)\citep{runcie_pitfalls_2019}.

	\subsection{Incremental comparison}
	We show how including secondary trait factors may help in increasing predictive ability (PA) by comparing three models which increase in model complexity and data used (Table \ref{comparison_table}).
	The first model is a univariate gBLUP using only marker data and grain yield.
	The second model is concatenated genetic latent factor BLUP (glfBLUP), where factors are constructed  over timeppoints using concatenated wavelength-timepoint  combinations as done in \citet{melsen_2025}.
	In addition, the correlation matrix is redundancy filtered and retained secondary traits are regularized.
	For a full explanation see \citep{melsen_2025}.
	The third model is glfBLUP as described in this study, using factors constructed at each timepoint, with and without Procrustes rotation.
	We compare the methods with secondary trait information available of incremental growth stage (e.g., from vegetative up to and including heading and grain filling) as indicated by \citep{krause_hyperspectral_2019}.
	We compare the PA of the above methods using the CIMMYT dataset for CV1 (with secondary traits only available in the training set) and CV2 (with secondary traits available in both training and test set) by splitting each subset of year-trial combination into 2/3 training and 1/3 test set, with 100 replications. 
	Furthermore, we demonstrate the interpretability of factors to show key timepoints and wavelengths for prediction of grain yield.
	\begin{center}
		\begin{table}[ht]
			\caption{Overview of models compared, growth stages of which data is available and whether Procrustes rotation is used. univariate gBLUP is used as benchmark based on markers and grain yield, so growth stages are not applicable}
			\begin{tabular}{c| c c c } 
				\hline
				Model & Growth stage(s) & Procrustes rotation \\ 
				\hline
				Univariate gBLUP & - & - \\
				Concatenated glfBLUP & vegetative & - \\ 
				Concatenated glfBLUP & vegetative, heading & - \\ 
				Concatenated glfBLUP & all & - \\ 
				Varimax factor per timepoint glfBLUP  & vegetative & no\\
				Varimax factor per timepoint glfBLUP & vegetative, heading & no \\
				Varimax factor per timepoint glfBLUP & all & no \\
				Procrustes factor per timepoint glfBLUP & vegetative  & yes \\
				Procrustes factor per timepoint glfBLUP & vegetative, heading & yes \\
				Procrustes factor per timepoint glfBLUP & all  & yes \\
				\hline
			\end{tabular}
			\label{comparison_table}
		\end{table}
	\end{center}
	
	\section{Results}
    We use a subset of the CIMMYT elite yield trial wheat dataset of 2014-2015 to demonstrate the application of factor analysis with Procrustes to model hyperspectral data collected over multiple timepoints.
    In addition, we use the extracted factor scores to relate the hyperspectral data to grain yield using subset selection and subsequently use selected factor scores in multivariate genomic prediction.
    We compare this to the predictive ability (PA) of univariate gblup and multivariate gblup using factor scores based on concatenated secondary traits (wavelength-timepoints combinations) as done in \citet{melsen_2025}.
    
	\subsection{Effects of drought treatment on grain yield}
\begin{figure}
		\centering
		\includegraphics[width=\textwidth]{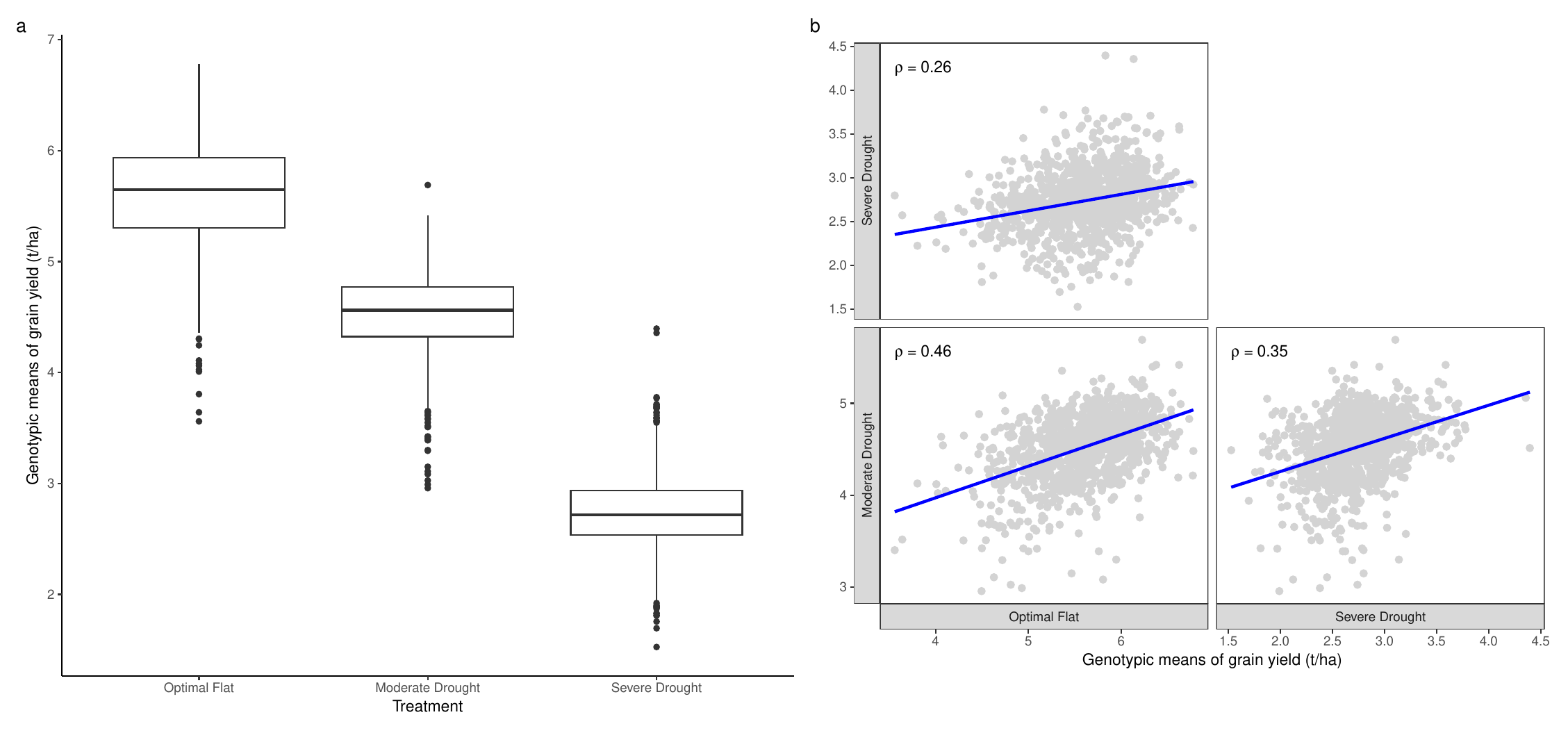}
		\caption{Boxplot (a) and scatter plots (b) of genotypic means (best linear unbiased estimators) of grain yield in tonnes per hectares (t/ha) indicated on the axis of the respective treatment within each scatter plot. Pearson correlation of each pair of treatments displayed as $\rho$.} 
		\label{fig:yield_pairsplot}
	\end{figure}

	We find a downward trend in grain yield with increasing drought conditions (Figure \ref{fig:yield_pairsplot} b), with optimal flat having the highest genotypic mean grain yield of 5.6 tonnes per hectare (t/ha) with a broad-sense heritability of 0.61 while this is 2.7 t/ha for the severe drought condition with a broad-sense heritability of 0.39.
	Similarly, the Pearson correlations based on genotypic means of grain yield decrease with stronger contrast in water treatments (Figure \ref{fig:yield_pairsplot} a).
	Here, the highest Pearson correlation (0.46) is observed between optimal flat and moderate drought treatments, while the lowest correlation (0.26) is observed between the optimal flat and severe drought treatments.
	\subsection{Interpretation of hyperspectral latent factors}
\begin{figure}[ht]
		\centering
		\includegraphics[width=.87\textwidth]{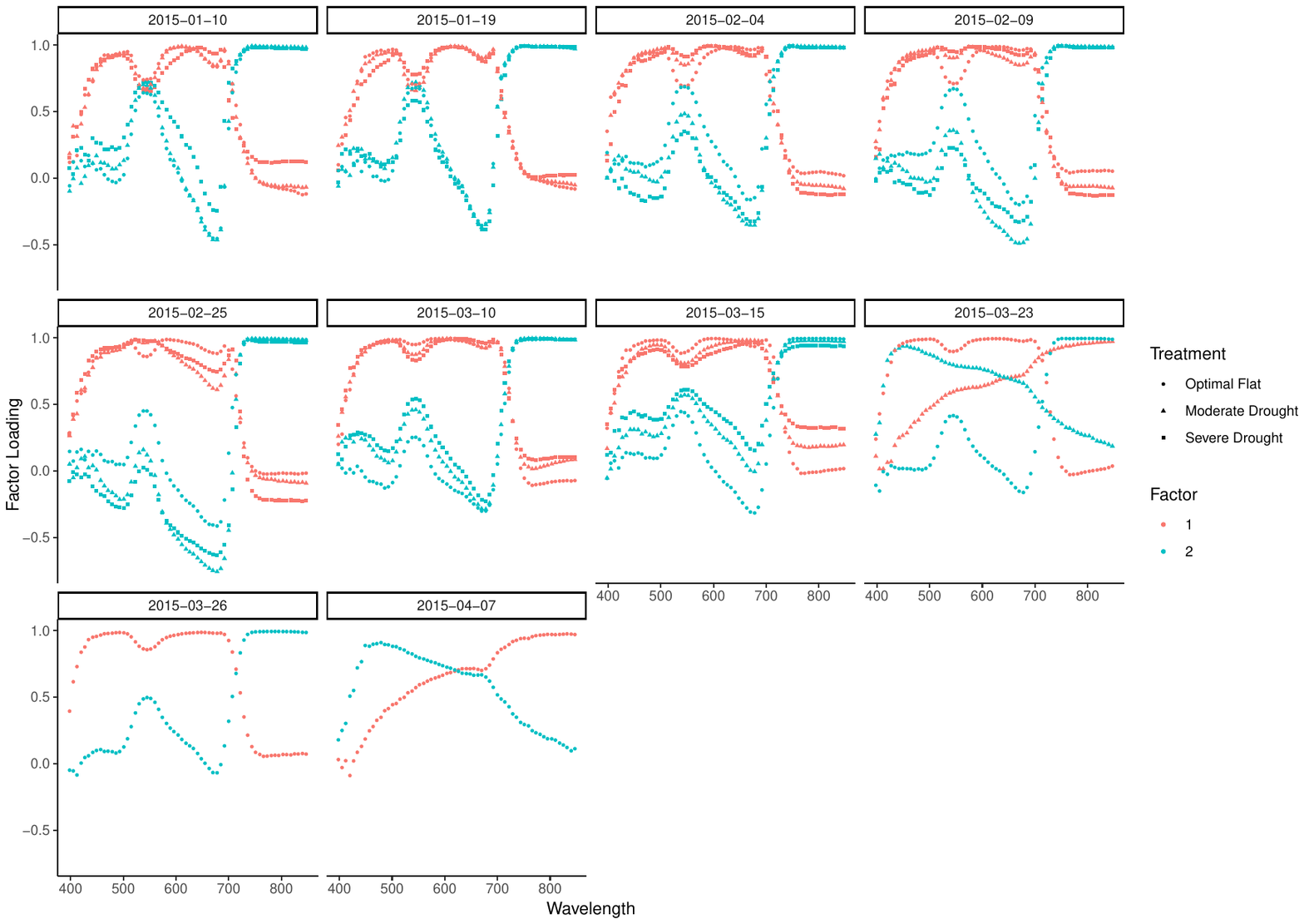}
        \\~\\
        \includegraphics[width=.87\textwidth]{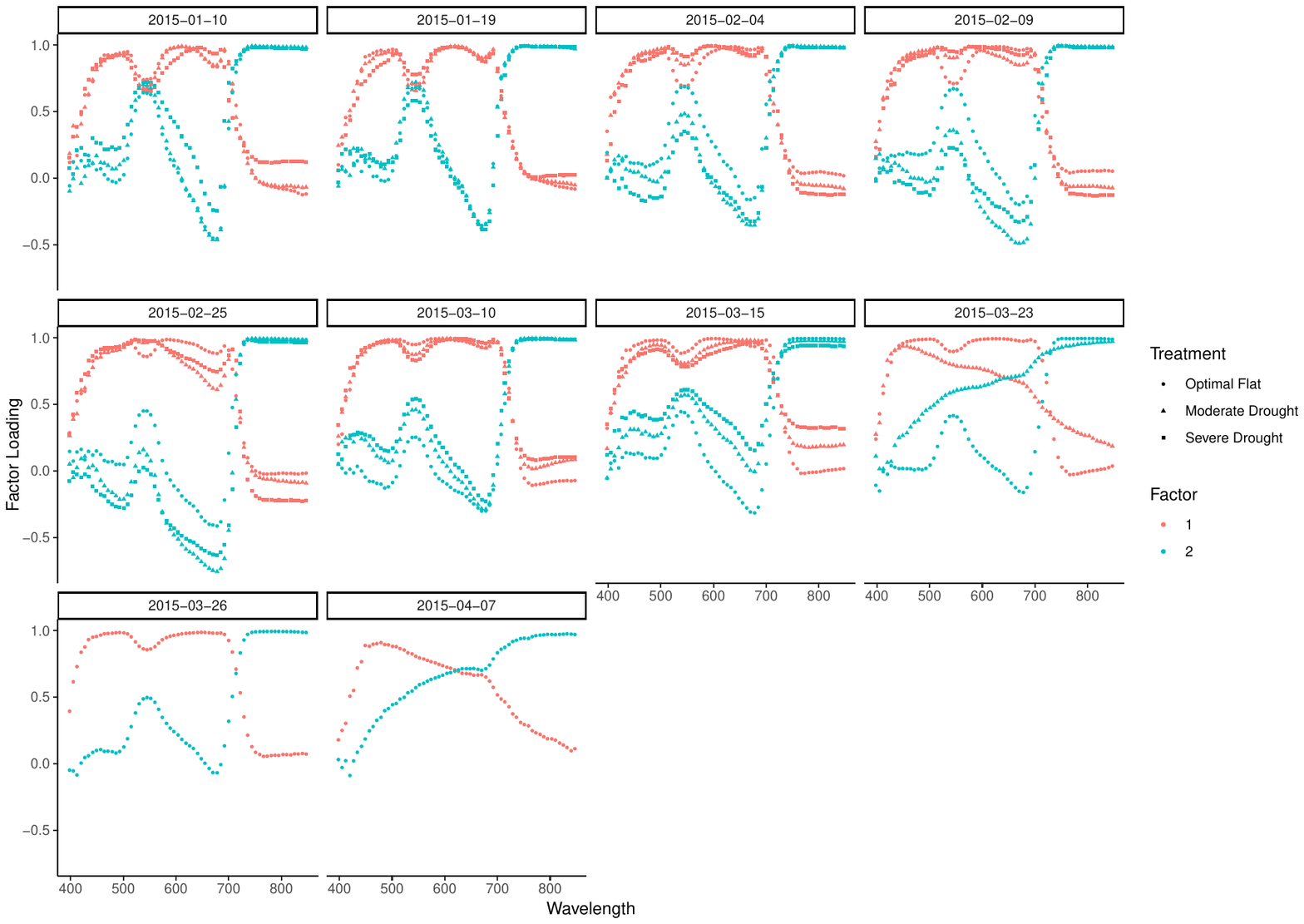}
		\caption{Wavelength reflectivity loadings on factor 1 and 2 of hyperspectral reflectivities for each timepoint and treatment before (top panel) and after (bottom panel) Procrustes rotation}
		\label{fig:loadings_plot}
	\end{figure}

\begin{figure}[ht]
		\centering
		\includegraphics[width=\textwidth]{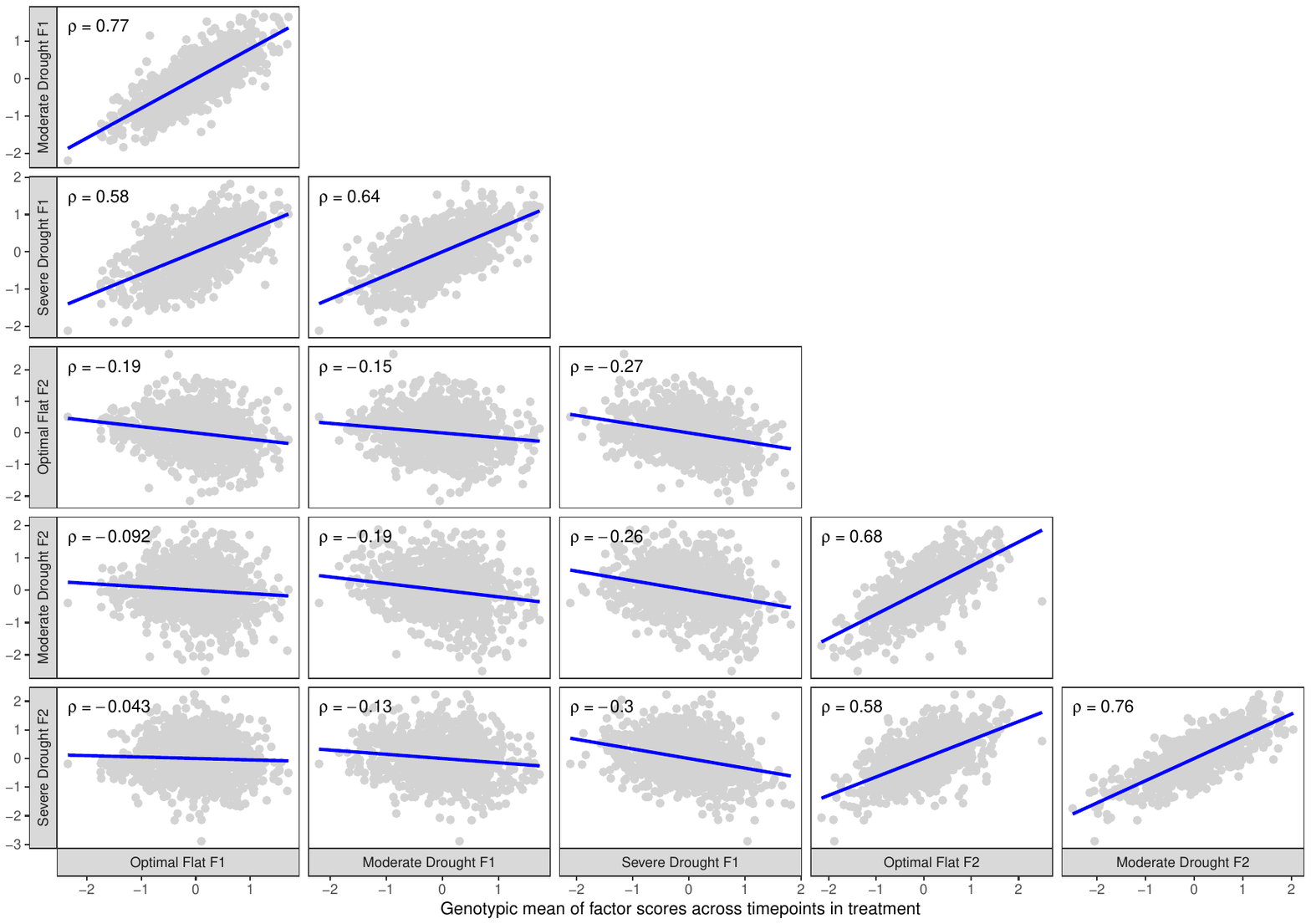}
		\caption{Scatter plots and Pearson's correlation ($\rho$) of genotypic mean of each factor score within each treatment. Factor is abbreviated in axis labels as 'F' for readability}
		\label{fig:factors_pairs}
	\end{figure}

	Before Procrustes rotation the loadings structure is already consistent across timepoints except for 2015-03-23 and 2015-04-07, where loadings switch between factors within both Optimal Flat and Moderate Drought treatments (Figure \ref{fig:loadings_plot}).
    More precisely, wavelengths within the range of 400 to 700 nm in timepoints before 2015-03-23 tend to have high loadings on factor 1 and low loadings on factor 2, while 700 to 850 nm tend to have high loadings on factor 2 and low loadings on factor 1. 
    This reverses after 2015-03-23.
	As expected, the Procrustes rotation and smoothing resolves this problem (Figure \ref{fig:loadings_plot}).
	In most timepoints a clear distinction is found between two factors, with factor 1 representing the spectrum between 400 to 700 nm, while factor 2 represents wavelengths above 700 nm. 
	Moreover, a small peak is consistently present in factor 2  around 560 nm across treatments. 
	Another similarity found is between moderate drought and optimal flat treatments with a change of factor loadings at the last timepoint, 23 March for moderate drought and 7 April for optimal flat.
	We find an increase in factor loadings in between 400 and 600 nm of factor 2. This change was not observed for the severe drought treatment, possibly due to early stopping of phenotyping because of drought. 
	As the factor structure remains relatively similar across treatments, we are able to correlate factor scores across the treatments (Figure \ref{fig:factors_pairs}). 
	As with the genotypic mean grain yield, the degree of correlation between the mean of factor scores between environments decreases as the treatments become less similar. 
    For example, the correlation for the mean of factor 1 for optimal flat and moderate drought is 0.74, while the correlation with severe drought is 0.54. The similarity between moderate and severe drought treatments becomes most clear within factor 2, with a correlation of 0.76.
    This reveals the distinct information retained within factor scores 1 and 2.
	\subsection{Subset selection}
\begin{figure}[ht]
		\centering
		\includegraphics[width=\textwidth]{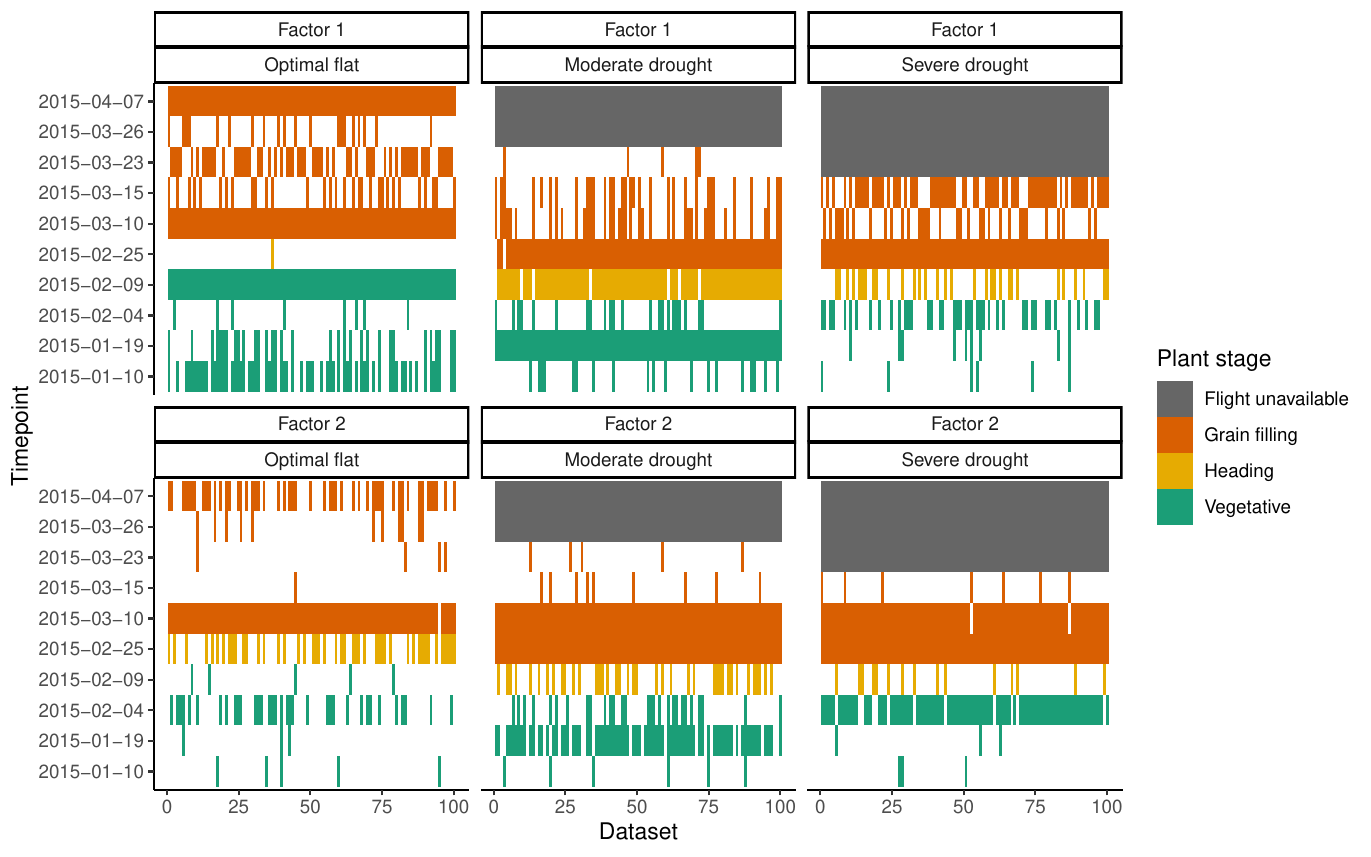}
		\caption{Factors included after subset selection for the CV2 scenario when all timepoints are available. A vertical dash indicating the timepoint factor being included in the final model for that training-test set. Timepoints in year-month-day format, plant stages according to \citet{krause_hyperspectral_2019}.
		Datasets created through cross-validation as used for predictive ability model comparison.}
		\label{fig:subset_selection}
	\end{figure}

	Subset selection was performed to relate the unsupervised factor scores to yield.
	The timepoint factors included differed across 100 datasets generated for cross validation. 
	However, a set of timepoint factors was consistently selected across datasets at 9 and 25 February, 10 March and 7 April (Figure \ref{fig:subset_selection}). 
	These dates correspond with timepoints succeeding the heading stage. 
	Surprisingly, the last timepoint was only consistently selected for factor 1 in the optimal flat treatment, while this was not the case for any of the factors in moderate and severe drought treatments. 
	This could indicate a difference in physiological response between environments such as an earlier start of senescence under drought conditions.
	Moreover, timepoints at heading stage were only included often for factor 1 in the moderate drought scenario.
	\subsection{Multivariate genomic prediction within different scenarios}
\begin{figure}[ht]
		\centering
		\includegraphics[width=\textwidth]{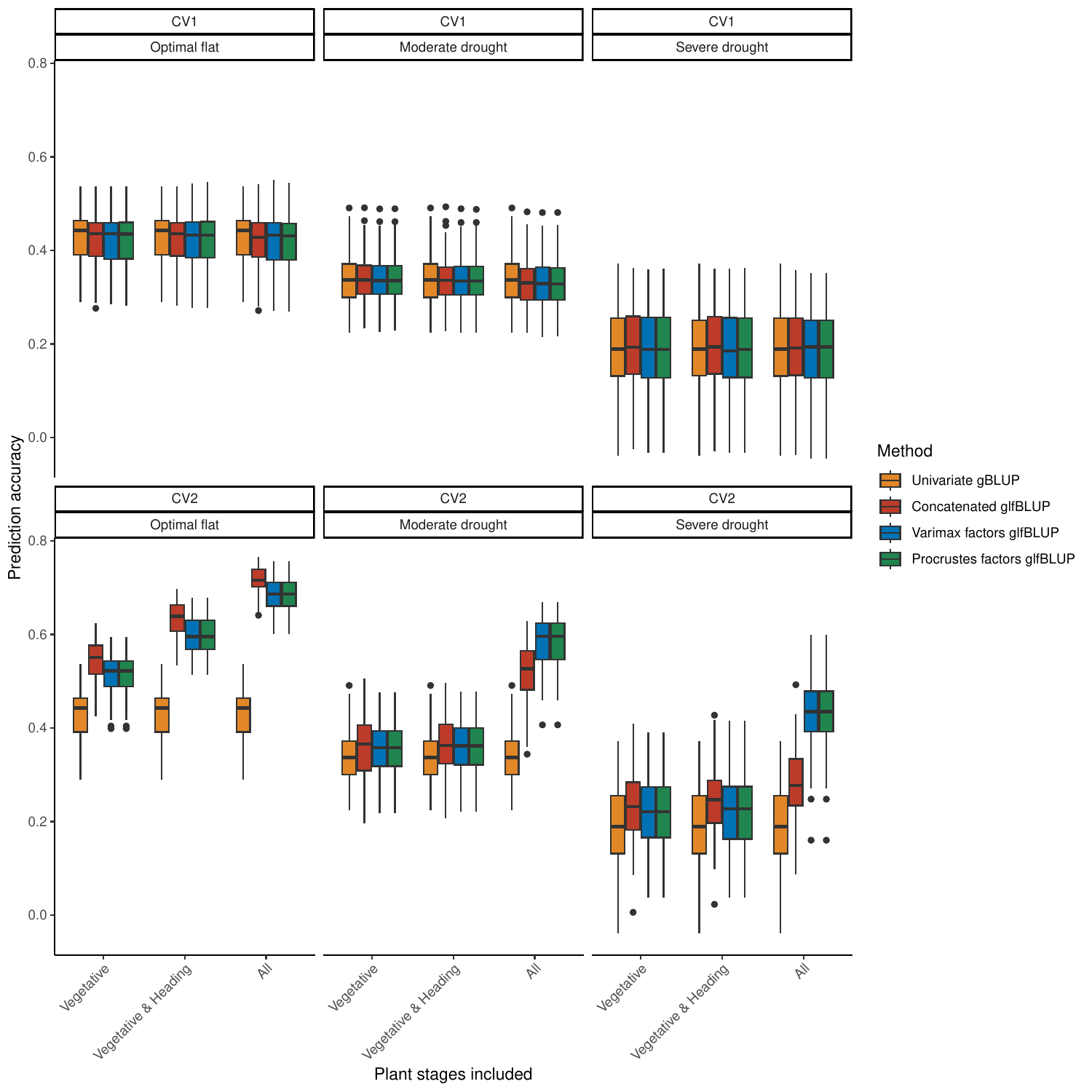}
		\caption{Pearson’s correlation between predicted and observed mean yield per genotype for each cross-validation (CV) scenario, treatment and plant stages considered per model type. Vegetative: only including secondary traits of vegetative stages. Vegetative \& Heading: Including secondary traits up to and including heading. All: Including secondary traits of all plant stages available.}
		\label{fig:accuracy}
	\end{figure}

	Improvement upon benchmark univariate gBLUP is found in all treatments within the cross-validation 2 (CV2) scenario (Figure \ref{fig:accuracy}). 
	Within this scenario, secondary traits are available within training and test set. 
    While glfBLUP with concatenated traits (wavelength-timepoint) combinations does outperform in the optimal flat treatment, this is not the case within both Moderate- and Severe drought.
	When all timepoints are considered, using factor scores per timepoint outperforms gBLUP and glfBLUP with concatenated traits in moderate and severe drought treatments. 
	However, this advantage is lost when only timepoints up to and including heading are available.
	In addition, splines were fit through Procrustes rotated factor scores and subsequently, the extracted spline characteristics where used in glfBLUP as secondary traits (Appendix\ref{sup:spline parameters}). 
	This did not lead to an increase in PA. 
    The additional required step of extracting spline characteristics possibly led to a loss of information. 
    However, smoothing splines can still be useful for visualization purposes of factor scores, because genotypes with a lower yield may have a different secondary trait response across the timepoints (Figure \ref{fig:factor_spline}).
	For the cross-validation 1 (CV1) scenario, where secondary traits are only used within the training set, all methods had a similar PA across treatments.

	\section{Discussion}
    High-dimensional phenotypic data is becoming more common within plant breeding research for arable crops. 
	Consequently, studies for integrating these data within genetic analysis and genomic prediction are increasingly required. 
	Here we show a computationally friendly way of incorporating time series hyperspectral phenotype data within multivariate genomic prediction for single trials. 
	While hyperspectral data is widely applied within plant sciences, we offer a general method that may be used for extracting interesting features using factor analysis, which can be traced across a plant growth cycle and retain interpretability.
    This is in contrast to current applications of factor analysis on phenotype data in which multiple timepoints are not yet involved  \citep{paixao_factor_2022, fialho_factor_2023} or difficult to interpret \citep{runcie_megalmm_2021}.
   
	Hyperspectral data regularly gets transformed through application of functions of specific wavelengths to extract vegetation indices such as the well established Normalized Difference Vegetation Index (NDVI) \citep{sellers_canopy_1985}. 
	Such indices are widely applied for proximal measurement of biomass related traits.
	However, selecting which indices to use or creating new indices is non-trivial with potential loss of information, as you utilize only fragments of the spectrum within the data deemed relevant.
	For example, \cite{krause_hyperspectral_2019} note that the blue region of the light spectrum may contain valuable plant physiological information as well, which is lost when using NDVI.
	In this study, we let the data decide by using a lower-dimensional representation of the hyperspectral data and relating these to the trait of interest (e.g., grain yield).
    We extract biologically relevant features by performing factor analysis on the genetic correlation matrix of hyperspectral time series data.
	Using Procrustes rotation, we were able to retain time series dynamics in a computationally efficient manner.
	Moreover, we demonstrate how these features are relatively stable across the season and treatments.
    Being of general nature, this method may reveal relevant features and patterns in other sensor data  not yet studied as extensively as hyperspectral data. 

    Although Procrustes rotation keeps the factors aligned across time, it does not guarantee that each factor reflects the same biology at every growth stage. 
    Wheat changes noticeably from vegetative growth to heading and grain filling, and its hyperspectral signals shift as well. 
    This is hinted at by the changes in factor structure we find in later timepoints. 
    A practical way to interpret the factors is to look at how they relate to familiar traits like NDVI \citep[see, e.g.,][]{melsen_2025}, canopy temperature, or chlorophyll, and how their key wavelengths evolve over time. 
    This can be accomplished through subset selection by changing the focal trait from grain yield to a familiar trait. 
    If these patterns follow known physiological stages—such as greenness peaking before heading or senescence under drought—then the factors are likely capturing meaningful biological changes.

	Performing factor analysis per timepoint provided us insight into which timepoints are important for prediction of grain yield in wheat. 
	Fitting the factor model per timepoint using all hyperspectral wavelength reflectivities could be done using standard estimation techniques, because the number of genotypes (1033) far exceeded the number of wavelengths (62).
    However, if the number of wavelengths exceeds the number of genotypes $s>g$, regularization steps can be taken to still fit a factor model, as done in \citet{melsen_2025}.
	We have shown that factor scores are unanimously selected around flowering time across all treatments. 
	While it is general knowledge that flowering time is an important indicator within wheat \citep{kole_genomics_2013}, we did not use this prior knowledge in the selection of timepoints. 
	Consequently, had we not known the importance of flowering time, it would have become evident from the analysis. 
	Here we used flowering time indication of the original author, where heading was indicated with 50 per cent of the plots being in the heading stage \citep{krause_hyperspectral_2019}.
	This range reinstates the importance for predicting flowering time, as it does shift across treatments while being an important indicator for grain yield.
	As traits like flowering time can be population dependent, robustness of variable selection should be studied within different populations.

	Quantitative genetic theory explains that a focal trait with lower heritability may be improved by selecting for a genetically correlated secondary trait with a higher heritability \citep{thompson_review_1986, runcie_pitfalls_2019}.
	This is a key objective for performing multivariate genomic prediction, where secondary trait information aids in prediction of a focal trait. 
	So far, this has mainly been proven to help in the cross-validation scenario 2 (CV2), where secondary trait information is available for both the training and test set genotypes. 
	However, how you incorporate secondary trait information determines the gain in predictive ability (PA). 
    One way is to concatenate timepoints and wavelengths before factor analysis \citep{melsen_2025}.
    Within the optimal treatment, we found multivariate prediction with factors based on such concatenated timepoint-wavelength traits to outperform multivariate prediction with timepoint factors.
	In contrast, in treatments with moderate and severe drought, multivariate genomic prediction with timepoint factors resulted in higher PA.
	The cause for this shift could be explained by subset selection being able to choose among relevant timepoints when factors are constructed per timepoint. 
	This is not the case for factors using concatenated timepoint-wavelength traits, as multiple timepoints can load on the same factor, making subset selection among timepoints difficult.

	In this study we used unsupervised factor analysis for dimensionality reduction and best subset selection for selecting timepoint factors to integrate high-dimensional phenotype data in multivariate genomic prediction.
	An alternative approach similar to \cite{arouisse_improving_2021} could be used. 
    In that case, all factor scores would be regressed on the focal trait and the subsequent phenotypic predictions would serve as input for a bivariate genomic prediction model.
	This would omit the requirement for best subset selection as done in this study.
	
	The frequency of flights and measurements within this dataset is rather low and at irregular intervals. 
	While this has no immediate effect on using factor scores, other methods for time series such a smoothing splines may be influenced. 
	In this study we used characteristics of the complete spline, which may have resulted in selected characteristics not providing additional PA (appendix 2). 
	Adding prior crop physiological knowledge might aid in extracting and selecting relevant developmental stages from said splines.

	As we show with the correlation analysis, secondary latent traits can help characterize treatments, opening the avenue towards multi-environment analysis. 
	Compared to correlational analysis across using yield data, factors based on hyperspectral data showed stronger correlations of treatments, resulting in potentially a better representation of genotypic response across treatments.
    In addition, the correlation between orthogonal factors (e.g., between factor 1 and factor 2), showed a near zero correlation, indicating both factors containing unique information.
    Besides such interpretations, the reduction of traits considered from 62 to 2 latent traits per treatment is beneficial in making analysis of larger multi-environment trials more computationally feasible. 
	Incorporation could then be done according to environmental kinships but other options in a factor analytic framework could be promising as well \citep{CostaNeto2020}.
	
	In conclusion, factor analysis is a flexible method for integrating and creating interpretable representations of high-dimensional secondary trait data for multivariate genomic prediction. 
	The results showcase the interpretability of secondary trait data and successful integration in multivariate genomic prediction.
	As such, this method can contribute to a greater understanding of high-dimensional data in plant breeding trials.

	\bigskip
	\subsection*{Data and code}
	Scripts to generate the hyperspectral datasets, as well as the results presented in this paper, are available  at \url{https://github.com/KunstJF/glfBLUP-Procrustes}. The glfBLUP methodology is implemented in an R-package available at \url{https://github.com/KillianMelsen/ glfBLUP}. The hyperspectral dataset is available upon reasonable request from J.\ Crossa	

	\bibliographystyle{apalike}
	\bibliography{gfBLUP_Applications_refs}
    \section{Statements and Declarations}
    \subsection*{Acknowledgements}
    We thank Margaret Krause and CIMMYT for generation and publication of the original data. 
    In addition, we thank the Graduate School for Production Ecology \& Resource Conservation (PE\&RC) for funding this study via the PE\&RC Graduate Programme call 2022 under number 22145.

    \subsection*{Funding}
	This study was funded by the Graduate School for Production Ecology \& Resource Conservation (PE\&RC).

    \subsection*{Competing Interests}
    The authors have no relevant financial or non-financial interests to disclose.

    \subsection*{Author Contributions}
    Conceptualization: C.F.W.P. and J.F.K; methodology and software: K.A.C.M. and J.F.K.; writing–original draft preparation: J.F.K; writing–review and editing: J.F.K., K.A.C.M., W.K., J.C., M.R.K., C.M., F.A.E., C.F.W.P. All authors have read and agreed to the published version of the manuscript.

	\clearpage



	\appendix
	\section{Factor dimension determination}
	\label{sup:dimension}
	When performing factor analysis the number of factors (i.e., dimension) needs to be determined. The upper-bound can determined by the acceleration factor \citep{raiche_non-graphical_2013}, an alternative to visual scree-plot elbow detection, where the dimension is at the point preceding highest acceleration. This means evaluating the second derivative of eigenvalue $\hat{\delta}_j$ from a linear regression using the optimal coordinate of the last eigenvalue and $(i + 1)^{th}$:
	\begin{equation*}
		\hat{\delta}_j = a_{j+1} + b_{j + 1} \left(j\right) \text{,}
	\end{equation*}
	where $a$ is the intercept and $b$ the slope with
	\begin{equation*}
		b_{j+1} = \frac{\delta_s - \delta_{j+1}}{s - j -1} \text{ and } a_{j+1} \equiv \delta_{j + 1}\left(j + 1\right) - b_{j+1}\left(j + 1\right) \text{.}
	\end{equation*}
	The second derivative is then approximated via:
	\begin{equation*}
		f''(j) = f(j + 1) - 2f(j) - f(j - 1)
	\end{equation*}
	where $f''(j)$ is the second order finite difference of eigenvalue ($\delta$) $j$. Meanwhile, we used the lower-bound Kaiser's rule where $k \ge 1$, resulting in the following choice $m^*$ for $m$ as described by \citep{raiche_non-graphical_2013}:
	\begin{equation*}
		m^* = \sum_{j} I(\delta_j \ge 1 \wedge j < s') \text{ with }s' \equiv \text{arg max}_j \left(af_j\right) \text{.}
	\end{equation*}
	\section{Extracting and using spline parameters}
	\label{sup:spline parameters}
\begin{figure}[ht]
		\centering
		\includegraphics[width=\textwidth]{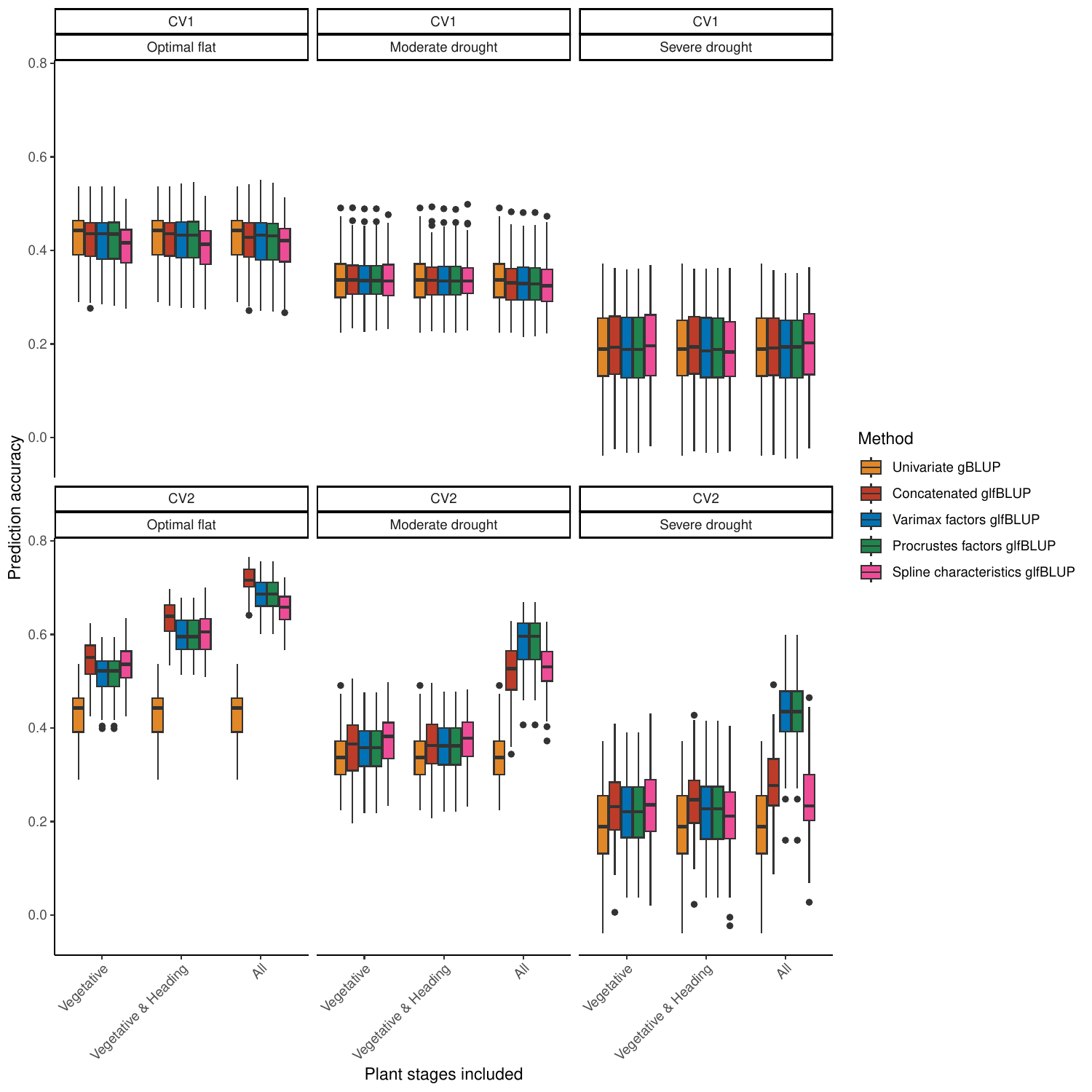}
		\caption{Pearson’s correlation between predicted and observed mean yield per genotype for each cross-validation (CV) scenario, treatment and plant stages considered per model type. Vegetative: only including secondary traits of vegetative stages. Vegetative \& Heading: Including secondary traits up to and including heading. All: Including secondary traits of all plant stages available.}
		\label{fig:accuracy_wspline}
	\end{figure}

	After estimation of the factor scores, we have reduced the dimension of $s$ secondary traits to $m^*$ factors, which are amenable to modelling across time through smoothing splines \cite{perez-valencia_two-stage_2022}.
	The population and genotype specific smoothing splines are fitted using the statgenHTP package, which is possible due to efficient calculation by the LMMsolver package \cite{boer_tensor_2023}.
	Each Procrustean smoothed factor score $\xih^l_{\pi ci}$ for plot $i$, genotype $c$ and population $\pi$ can be modeled across timepoints $l$ with a hierarchical smoothing spline:
	\begin{equation*}
		\xih_{\pi ci}\left(l\right) = \phic_\pi\left(l\right) + \phic_{\pi c}\left(l\right) + \phic_{\pi ci}\left(l\right) + \epsilon_{\pi ci}\left(l\right), \quad \epsilon_{\pi ci}\left(l\right)\sim \mathcal{N}_n\left(0, \sigma^{2}\left(l\right)\right),
	\end{equation*}
	with the population $\pi$ spline being $\phic_\pi\left(l\right)$, the genotype $c$ specific spline $\phic_{\pi c}\left(l\right)$ and the plot $i$ specific spline being $\phic_{\pi ci}\left(l\right)$.
	The genotype specific splines can be seen as deviations from the population spline.
	To use these splines in aiding prediction of yield, key parameters of the genotypic spline are extracted per factor. 
	For example, the area under the curve of the interval between the vegetative and grain filling stages, as this is a moment when large genotypic variance may be observed. 
	Determination of other intervals for parameter extraction can be done by visual inspection of the genotype-specific splines. 
	These characteristics are subsequently used as secondary traits in genomic prediction.
	However, there was no observed benefit of  substituting factor scores with spline characteristics as secondary traits in the multivariate gblup equations (Figure \ref{fig:accuracy_wspline}).
	\clearpage
	\section{Univariate and multivariate BLUP equations}
	\label{sup:blup_equations}
    We present general univariate and multivariate BLUP equations, where $f$ denotes the focal trait, $w$ the secondary trait(s),  subscript $o$ the training set (e.g., observed phenotypes) and subscript $u$ the test set (e.g., unobserved phenotypes). So, we denote the training-test split phenotype matrix as:
    \begin{equation*}
        \mathbf{Y} = \begin{bmatrix}
             \mathbf{Y}_o \\
            \mathbf{Y}_u
        \end{bmatrix} .
    \end{equation*}
	First we introduce univariate gBLUP on a focal trait $\y^f$ and then extend it to multivariate gBLUP for both CV1 and CV2.
	For univariate gBLUP we use the following notation:
	\begin{equation*}
		\begin{split}
			\Vh&=\hat{\sigma}^f_{G}\otimes \K_{o}+\hat{\sigma}^f_{E}\otimes \I_{n} \\
			\gh^{(uni)}_{fo}&=(\hat{\sigma}^f_{G}\otimes \K_{o})\Vh^{-1}\big(\y^f_{o}-\X_{o}\hat{\boldsymbol{\beta}}^f]\big) \\
			\gh^{(uni)}_{fu}&=\mathbb{E}(\g_{fu}|\y_{fo}) \\
			&=(\hat{\sigma}^f_{G}\otimes \K_{uo})\Vh^{-1}\big(\y^f_{o}-\X_{o}\hat{\boldsymbol{\beta}}^f\big) \\
			&=\K_{uo}\K_{o}^{-1}\gh^{(uni)}_{fo} \text{,}
		\end{split}
	\end{equation*}
	where $\gh^{(uni)}_{fo}$ and $\gh^{(uni)}_{fu}$ are the focal trait univariate BLUPs for the training and test set.
	Also note that as $\hat{\sigma}^f_{G}$ and $\hat{\sigma}^f_{E}$ are scalars, $\Vh=\hat{\sigma}_{Gf} \K_{o}+\hat{\sigma}_{Ef}\I_{n}$.
	We extend the univariate gBLUP model to a multivariate version. Here we use the genetic ($\Covmat_{G}$) and residual ($\Covmat_{E}$) covariance matrices estimated from the data of the focal ($f$) and secondary traits ($w$) (i.e., factor scores):
    	\begin{equation*}
        \begin{split}
        \Covmat_{G} &= \begin{bmatrix}
			\Covmat^{w}_G  \quad \boldsymbol{\sigma}^{w f}_G\\
			\boldsymbol{\sigma}^{fw}_G \quad \sigma^{f}_G
		\end{bmatrix}  = \begin{bmatrix}
		\boldsymbol{\sigma}^{w.}_G \\
        \boldsymbol{\sigma}^{f.}_G
		\end{bmatrix} \\
		\Covmat_{E} &= \begin{bmatrix}
			\Covmat^{w}_E \quad \boldsymbol{\sigma}^{w f}_E\\
			\boldsymbol{\sigma}^{f w}_E \quad \sigma^{f}_E
		\end{bmatrix} = \begin{bmatrix}
		\boldsymbol{\sigma}^{w.}_E \\
        \boldsymbol{\sigma}^{f.}_E
		\end{bmatrix} \text{.}
        \end{split}
	\end{equation*} 
    In the case of CV1 (secondary traits are unobserved in the test set):
	\begin{equation*}\label{CV1_preds}
		\begin{split}
			\Vh&=\Covmath_{G}\otimes \K_{o}+\Covmath_{E}\otimes \I_{n} \\
			\gh^{(CV1)}_{fo}&=(\boldsymbol{\sigma}^{f.}_G\otimes \K_{o})\Vh^{-1}\vect\big(\Y_o-\X_o\hat{\boldsymbol{\beta}}\big) \\
			\gh^{(CV1)}_{fu}&=\mathbb{E}(\g_{fu}|\Y_o) \\
			&=\big(\boldsymbol{\sigma}^{f.}_G\otimes \K_{uo}\big)\Vh^{-1}\vect\big(\Yh_o-\X_o\hat{\boldsymbol{\beta}}\big) \\
			&=\K_{uo}\K_{o}^{-1}\gh^{(CV1)}_{fo} \text{,}
		\end{split}
	\end{equation*}
	where $\Y_{o}-\X_o\hat{\boldsymbol{\beta}}$ is vectorized.
	The multivariate predictions of the CV2 scenario (secondary traits are observed in the test set), can be obtained directly \citep{arouisse_improving_2021} or through a two-step approach \citep{runcie_pitfalls_2019}.
	We prefer the two-step approach because the algorithm can be used to obtain all CV1 training set BLUPs, $\Gh^{(CV1)}_o$:
	\begin{equation*}\label{CV2_preds_1}
		\begin{split}
			\Vh&=\Covmath_{G}\otimes \K_{o}+\Covmath_{E}\otimes \I_{n} \\
			\Gh^{(CV1)}_{o}&=\vect^{-1}(\Covmath_{G}\otimes \K_{o})\Vh^{-1}\vect(\Y_{o}-\X_o\hat{\boldsymbol{\beta}}) \text{.}
		\end{split}
	\end{equation*}
	We then convert these training set BLUPs to intermediate CV1 test set BLUPs and combine those with the observed secondary features for the test set to finally obtain CV2 BLUPs for the test set focal trait:
	\begin{equation*}\label{CV2_preds_2}
		\begin{split}
			\Vh&=\Covmath_{G}\otimes \K_{u}^{-1}+\Covmath_{E}\otimes \I_{n} \\
			\gh^{(CV2)}_{fu}&=\mathbb{E}(\g_{fu}|\Y_{wu},\Y_o) \\
			&=\K_{uo}\K_{o}^{-1}\gh^{(CV1)}_{fo}+(\boldsymbol{\sigma}^{wf}\otimes \K_{u}^{-1})\Vh^{-1}\vect(\Y_{wu}-\K_{uo}\K_{o}^{-1}\Gh^{(CV1)}_{o}) \text{.}
		\end{split}
	\end{equation*}	

\end{document}